\documentclass{webofc}\usepackage[varg]{txfonts}
\usepackage{epsfig,graphics,amssymb,amsmath,mathrsfs,color,colortbl,
bm,booktabs,cool,dcolumn}

\begin{document}\title{Zooming in on Multiquark Hadrons\\within QCD
Sum-Rule Approaches}\author{Wolfgang Lucha\inst{1}\fnsep\thanks
{\email{Wolfgang.Lucha@oeaw.ac.at}}\and Dmitri Melikhov
\inst{2,3,4}\fnsep\thanks{\email{dmitri_melikhov@gmx.de}}\and
Hagop Sazdjian\inst{5}\fnsep\thanks
{\email{sazdjian@ijclab.in2p3.fr}}}\institute{Institute for High
Energy Physics, Austrian Academy of Sciences, Nikolsdorfergasse
18,\\A-1050 Vienna, Austria\and D.~V.~Skobeltsyn Institute of
Nuclear Physics, M.~V.~Lomonosov Moscow State University,\\119991
Moscow, Russia\and Joint Institute for Nuclear Research, 141980
Dubna, Russia\and Faculty of Physics, University of Vienna,
Boltzmanngasse 5, A-1090 Vienna, Austria\and Universit\'e
Paris-Saclay, CNRS/IN2P3, IJCLab, 91405 Orsay, France}
\abstract{Aiming at self-consistent descriptions of multiquark
hadrons (such as tetraquarks, pentaquarks, hexaquarks) by means of
QCD sum rules, we note that the totality of contributions to
two-point or three-point correlation functions that involve,
respectively, either two or just a single operator capable of
interpolating the particular multiquark under study can be
straightforwardly disentangled into two disjoint classes defined
by unambiguously identifiable members. The first is formed by
so-called multiquark-phile contributions which indeed might
support multiquarks. In the case of flavour-exotic tetraquarks, by
definition composed of four (anti-) quarks of mutually different
flavours, a tetraquark-phile contribution has to exhibit two or
more gluon exchanges of appropriate topology. The second consists
of contributions evidently not bearing any relation to
multiquarks; these must be discarded when studying multiquarks by
QCD sum rules. The first class only should enter the
``multiquark-adequate'' QCD sum rules for exotic hadrons.}
\maketitle

\section{Multiquark Hadrons, Subset of Unconventional QCD Bound
States}\label{M}The relativistic quantum field theory that
provides the fundamental-level description of strong interactions
in terms of quarks and gluons acting as fundamental degrees of
freedom, quantum chromodynamics (QCD), in principle permits, as
bound states of these degrees of freedom, all combinations that
form a singlet under its non-Abelian gauge group SU(3). Therefore,
among its bound states there have been observed not only the two
categories of \emph{conventional\/} hadrons, quark--antiquark
mesons and three-quark baryons, but even (candidates for)
representatives of various, likewise QCD-compatible categories of
so-called nonconventional or \emph{exotic\/} hadrons. To these
belong \emph{multiquark\/} hadrons (tetraquarks, pentaquarks,
hexaquarks, heptaquarks, \dots), quark--gluon bound states, called
hybrid, and pure-gluon bound states, (nick)named glueballs.

Within the framework of quantum field theory, the concept of QCD
sum rules \cite{QSR} provides a nonperturbative analytic (that is,
not merely numerical) approach to bound states. QCD sum rules
represent a means to \emph{comparatively\/} easily trace back
observable properties of hadrons to all the degrees of freedom and
parameters entering the underlying quantum field theory, QCD.
Quite generally, their derivation proceeds along a sequence of
manipulations the main steps of which may be outlined by the
following recipe: Identify or, if necessary, construct an operator
that interpolates the hadron in your focus of interest, by having
a nonvanishing matrix element if sandwiched between the vacuum and
the particular hadron state, from the pool of quark and gluon
field operators offered by QCD. Consider an apparently convenient
correlation function of hadron interpolating operators. Evaluate
the correlation function at both phenomenological hadron and
fundamental QCD level. To this end: Apply the operator~product
expansion \cite{KGW}, in order to separate perturbative from
nonperturbative contributions. Allow the hadron spectrum to enter
the game by inserting a complete set of hadron states. Emphasize
lowest-mass hadron states by utilizing fitting Borel
transformations. Assume that, above (ideally optimized
\cite{ET1,ET2,ET3}) effective thresholds, perturbative-QCD and
hardly known higher-hadron contributions cancel.

Our present intention is to lift the standard QCD sum-rule
formalism encoded in the~above prescription -- in particular, the
theoretical starting point, viz., the QCD input to the correlation
function -- to a level better matching the group-theoretically
induced peculiarities of the exotic multiquark hadrons that
strikingly distinguish them from the class of ordinary hadrons
\mbox{\cite{ESRp,ESRr,TMA1,TMA2,LMS10,TMA3}}. Specifically, for
well-defined categories of tetraquarks \cite{LMS4}, we are led to
suspect that this goal necessitates a QCD contribution to involve
two or more gluon exchanges of suitable~topology.

Below, we recall in brief some of our earlier considerations
\cite{ESRp,ESRr,LMS10} on how to adequately optimize the QCD
sum-rule approach to multiquark states. In this context, the main
challenge is to identify and subsequently single out precisely
those contributions to correlation functions that have the
potential to influence QCD sum-rule predictions of multiquark
features (Sect.~\ref{T}). Upon following the route materializing
thereby, anyone's quest for increased precision should be rewarded
by eventually arriving at so-called multiquark-adequate QCD sum
rules~(Sect.~\ref{S}), as illustrated, for clarity, for the
tetraquarks with ``flavour-exotic'' quark composition
(Sect.~\ref{E}). Ignoring all for the subsequent line of argument
irrelevant issues (like parity or spin degrees of freedom, by
notationally suppressing Dirac matrices) will enable us to focus
on the essentials.

\section{Tetraquark Mesons, Conceptually Simplest Form of Exotic
Hadron}\label{T}At present, the -- at least from the experimental
perspective \cite{PDG} -- presumably best established category of
exotic multiquark hadrons seems to be formed by the totality of
tetraquark mesons\begin{equation}T=[\overline q_a\,q_b\,\overline
q_c\,q_d]\ ,\label{tm}\end{equation}bound states of two quarks
$q_b,q_d$ and two antiquarks $\overline q_a,\overline q_c$
carrying flavour quantum numbers\begin{equation}a,b,c,d\in
\{u,d,s,c,b\} \label{qn}\end{equation}and having masses
$m_a,m_b,m_c,m_d$ that (like the strong coupling) are \emph{basic
parameters\/} of~QCD and will assume a prominent r\^ole in the
tetraquark-adequate QCD sum rules drafted in Sect.~\ref{S}.

Following, for the intended construction of the appropriate
tetraquark-adequate QCD sum rules, the prescription recapitulated
above, we first have to specify the tetraquark interpolating
operators we would like to employ. Our task is significantly
facilitated by the observation \cite{RLJ} that, by application of
suitable Fierz transformations \cite{MF}, any chosen tetraquark
interpolating operator may be demonstrated to be equivalent to a
linear combination of merely two products\begin{equation}
\theta_{\overline ab\overline cd}(x)\equiv j_{\overline ab}(x)\,
j_{\overline cd}(x)\ ,\qquad\theta_{\overline ad\overline cb}(x)
\equiv j_{\overline ad}(x)\,j_{\overline cb}(x)\label{t}
\end{equation}of SU(3)-singlet quark--antiquark bilinear operators
$j_{\overline ab}(x)$ interpolating conventional mesons,
\begin{equation}j_{\overline ab}(x)\equiv\overline q_a(x)\,q_b(x)\
.\label{b}\end{equation}(Useful operator identities are provided
by Ref.~\cite{LMS111}, Eqs.~(32), (36), or Ref.~\cite{ESRr},
Eqs.~(1),~(2).)\clearpage

Since having in mind to unveil the deepest secrets of the
tetraquark mesons by scrutinizing their effects upon contributing,
by way of intermediate-state poles, to amplitudes encoding~the
scattering of two conventional mesons into two conventional
mesons, a smart idea is to trust in correlation functions of a
time-ordered product of four quark--antiquark bilinear
operators~(\ref{b}),\begin{equation}\left\langle{\rm T}\!\left(
j(y)\,j(y')\,j^\dag(x)\,j^\dag(x')\right)\right\rangle.\label{4}
\end{equation}Configuration-space contractions of both pairs of
quark--antiquark bilinear operators (\ref{b}) yield correlation
functions of two tetraquark interpolating operators (\ref{t})
(cf.\ Subsect.~\ref{p2.} or \ref{r2.}):\begin{equation}\left
\langle{\rm T}\!\left(\theta(y)\,\theta^\dag(x)\right)\right
\rangle=\lim_{\underset{\scriptstyle y'\to y}{\scriptstyle x'\to
x}}\left\langle{\rm T}\!\left(j(y)\,j(y')\,j^\dag(x)\,j^\dag(x')
\right)\right\rangle.\label{2}\end{equation}Configuration-space
contractions of one pair of quark-bilinear operators (\ref{b})
create correlation functions involving only one tetraquark
interpolating operator (\ref{t}) (cf.\ Subsect.~\ref{p3.} or
\ref{r3.}):\begin{equation}\left\langle{\rm T}\!\left(j(y)\,j(y')
\,\theta^\dag(x)\right)\right\rangle=\lim_{x'\to x}\left\langle
{\rm T}\!\left(j(y)\,j(y')\,j^\dag(x)\,j^\dag(x')\right)\right
\rangle.\label{3}\end{equation}

Now, from our point of view, at least, the decisive move towards
the envisaged multiquark adequacy \cite{TMA2} of QCD sum rules is
the unambiguous sorting out of all QCD-level contributions that
promise to bear potential relevance for multiquarks from those
that, beyond doubt, do not. Accordingly, in order to sharpen our
blades, we formulated, for the case of tetraquark~mesons, a simple
and, in fact, rather self-evident criterion \cite{TQC1,TQC2} that
enables us to separate the~wheat from the chaff by identifying
those contributions, henceforth named \emph{tetraquark-phile\/}
\cite{TQP1,TQP2}, that may contribute to the expected formation of
some tetraquark pole: Let $p_1,p_2$ and $q_1,q_2$ be the four
external momenta related to a generic correlation function
(\ref{4}). Recall the definition of the Mandelstam variable $s$ in
terms of initial and final momenta $p_1,p_2$ and $q_1,q_2$,
respectively,\begin{equation}s\equiv(p_1+p_2)^2=(q_1+q_2)^2\
.\end{equation}We deem a QCD-level contribution, considered as
function of $s$, \emph{tetraquark-phile\/} if it exhibits a
nonpolynomial dependence on $s$ and if it develops a branch cut
that starts at the branch point\begin{equation}\hat s\equiv
(m_a+m_b+m_c+m_d)^2\ .\end{equation}Whether or not the requirement
of the existence of an appropriate branch point is fulfilled~may
easily be decided by inspection of the correlation function in
question by means of the Landau equations \cite{LDL}. (An
illustration of their application may be found in the Appendix of
Ref.~\cite{ESRr}.)

\section{Tetraquark Mesons: Genuinely Flavour-Exotic Quark
Composition}\label{E}In order to keep all following applications
of our notion of multiquark adequacy as transparent as possible,
let us focus to the \emph{flavour-exotic\/} tetraquarks: bound
states of four (anti-) quarks of mutually different quark
flavours, specified by further restriction of the
characterization~(\ref{qn})~by\begin{equation}T=[\overline
q_a\,q_b\,\overline q_c\,q_d]\ ,\qquad a,b,c,d\in\{u,d,s,c,b\}\
,\qquad a\ne b\ne c\ne d\ .\label{F}\end{equation}

Resorting to our criterion of Sect.~\ref{T}, we identify QCD-level
contributions to the four-point correlation functions (\ref{4})
maybe decisive for tetraquarks (\ref{F}) by sorting out any
irrelevant one. For these, the configuration-space contractions
(\ref{2}) and (\ref{3}) provide the sought tetraquark-phile
two-point and three-point correlation functions; pairing any such
correlation function with the appropriate hadron-level counterpart
then yields the corresponding QCD sum rule, cf.\ Sect.~\ref{S}.

For (at least) flavour-exotic tetraquarks, the correlation
functions (\ref{4}) may be discriminated according to whether
quark-flavour distributions of incoming and outgoing states are
identical (Subsect.~\ref{p}) or different (Subsect.~\ref{r}); we
tag the latter possibilities ``flavour-preserving'' or
``flavour-retaining'' and ``flavour-rearranging'' or
``flavour-reordering'' or the like, respectively.

\subsection{Flavour-Preserving Four-Point Correlation Functions of
Interpolating Currents}\label{p}With respect to the complexity of
the analysis, this task is definitely easier if the distribution
of the quark flavours is identical in initial and final state of
the four-point correlation function~(\ref{4}),\begin{equation}
\left\langle{\rm T}\!\left(j_{\overline ab}(y)\,j_{\overline cd}
(y')\,j^\dag_{\overline ab}(x)\,j^\dag_{\overline cd}(x')\right)
\right\rangle,\qquad\left\langle{\rm T}\!\left(j_{\overline ad}(y)
\,j_{\overline cb}(y')\,j^\dag_{\overline ad}(x)\,
j^\dag_{\overline cb}(x')\right)\right\rangle.\label{fp}
\end{equation}

\subsubsection{Two-Point Correlation Functions: Two Identical
Tetraquark Interpolating Operators}\label{p2.}
\begin{figure}[htb]\centering{\includegraphics[scale=.401,clip]
{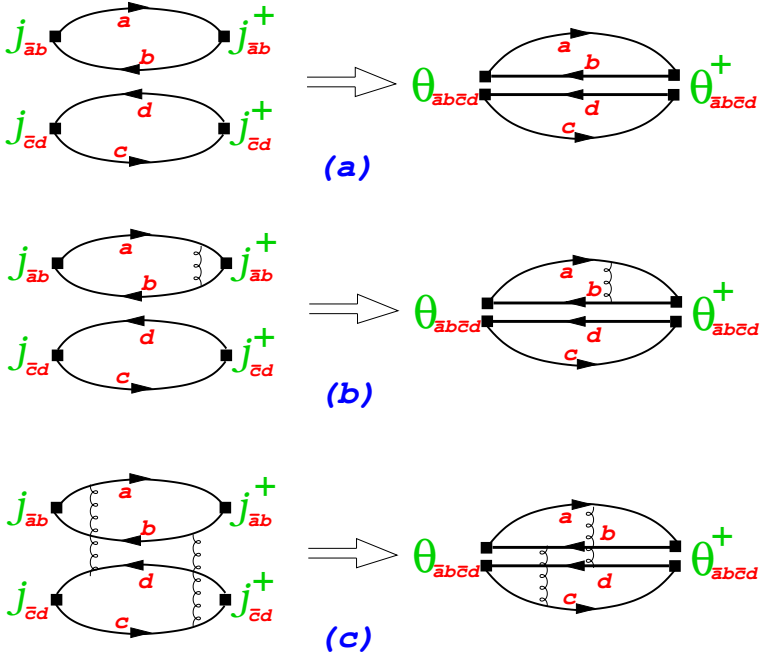}}\caption{Flavour-\emph{preserving\/}
correlation function \cite{ESRp,TMA2}: representative examples of
contributions of the three (a,b,c) lowest perturbative orders and
their contractions to two tetraquark-interpolating
operators~$\theta$.}\label{Fp2}\end{figure}

\begin{figure}[htb]\centering{\includegraphics[scale=.27765,clip]
{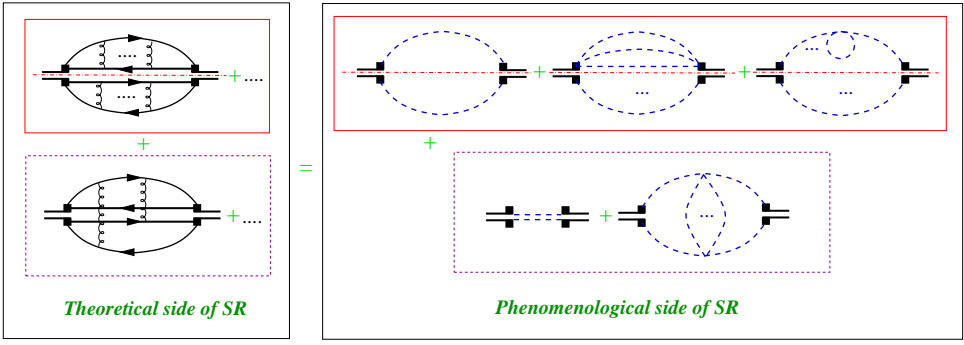}}\caption{Flavour-\emph{preserving\/}
two-tetraquark correlation function \cite{ESRp,TMA2},
disintegrable into (top row) two \emph{ordinary\/}-meson QCD sum
rules and (bottom row) a QCD sum rule possibly supporting
tetraquark~poles.}\label{Fsg}\end{figure}

\begin{figure}[htb]\centering{\includegraphics[scale=.50756,clip]
{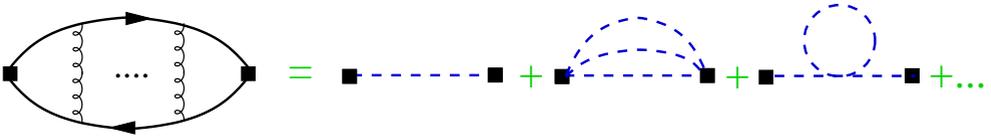}}\caption{Conventional QCD sum rule \cite{QSR}
preferably applicable to \emph{ordinary\/} mesons (blue dashed
lines).}\label{Fso}\end{figure}

\begin{figure}[htb]\centering{\includegraphics[scale=.47736,clip]
{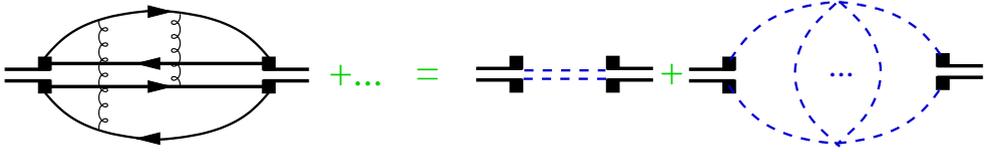}}\caption{Refined QCD sum rule \cite{ESRp}
likely improving analyses of \emph{tetraquarks\/} (blue dashed
double~line).}\label{Fst}\end{figure}

The analysis of the
flavour-preserving case is comparatively easy \cite{ESRp}: Figure
\ref{Fp2} exemplifies the lower-order perturbative QCD-level
contributions to the four-point correlation functions (\ref{fp}).
It poses no problem to convince oneself that any tetraquark-phile
among these features at least two gluon exchanges exhibiting
appropriate topology, like the one depicted in Fig.~\ref{Fp2}(c).
Then, as indicated by Fig.~\ref{Fsg}, the contraction (\ref{2}) of
a correlation function (\ref{fp}) decomposes into a pair of
ordinary-meson QCD sum rules (Fig.~\ref{Fso}), and a
tetraquark-friendlier QCD sum~rule~(Fig.~\ref{Fst}).

Phrased with a little bit sense of humour, the outcomes of the
aforegoing discussion for the flavour-preserving partition can be
subsumed by the kind of ``graphical-mathematics''~relation
\begin{equation}\mbox{Fig.~\ref{Fsg}}=2\times(\mbox{Fig.~\ref{Fso}})
+\mbox{Fig.~\ref{Fst}}\ .\label{gm}\end{equation}

\subsubsection{Three-Point Correlation Function -- One Tetraquark
and Two Conventional Mesons}\label{p3.}
\begin{figure}[htb]\centering{\includegraphics[scale=.401,clip]
{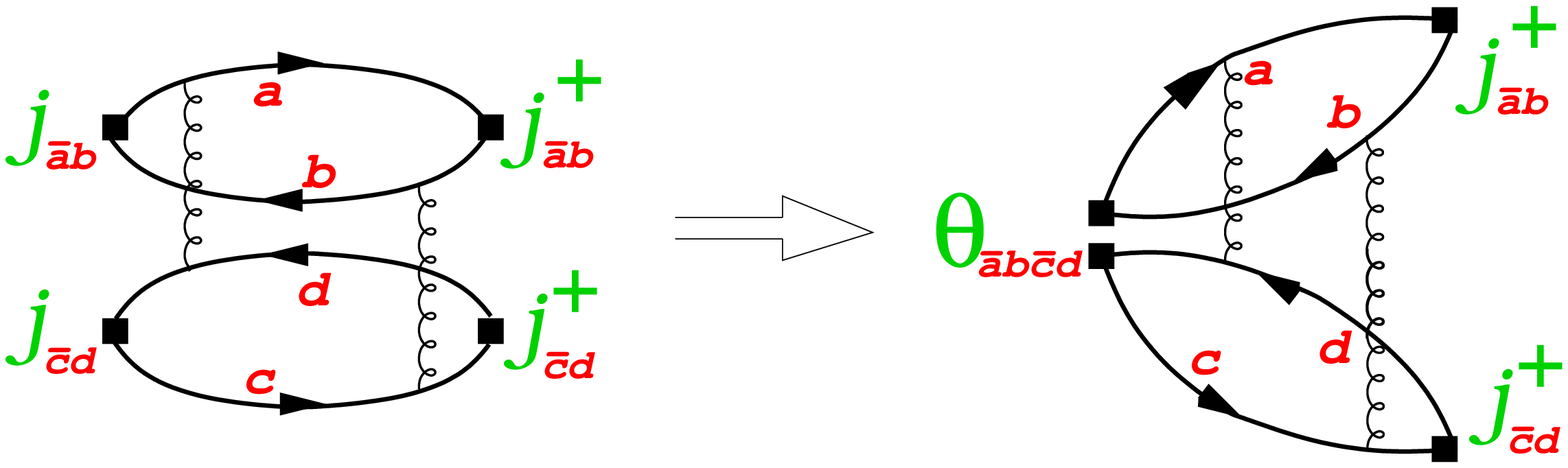}}\caption{Flavour-\emph{retaining\/}
correlation function \cite{ESRp}: typical lowest-order
\emph{tetraquark-phile\/} contribution, subject to a single
contraction to a tetraquark-interpolating operator, leaving two
quark-bilinear currents.}\label{Fp3}\end{figure}

Given the insights
established generally for the four-point correlation functions
(\ref{fp}), identical conclusions about nature of tetraquark-phile
contributions (Fig.~\ref{Fp3}), decomposition of deduced
quark--hadron relations and tetraquark-fitting QCD sum rules will
hold for the contraction (\ref{3}).

\subsection{Flavour-Regrouping Four-Point Correlation Function of
Interpolating Currents}\label{r}Easy to guess, we now turn to the
only other option for quark-flavour distribution in four-point
correlation functions (\ref{4}), viz., that with different flavour
arrangement in initial and~final~states:\begin{equation}\left
\langle{\rm T}\!\left(j_{\overline ab}(y)\,j_{\overline cd}(y')\,
j^\dag_{\overline ad}(x)\,j^\dag_{\overline cb}(x')\right)\right
\rangle.\end{equation}

\begin{figure}[h]\centering{\includegraphics[scale=.401,clip]
{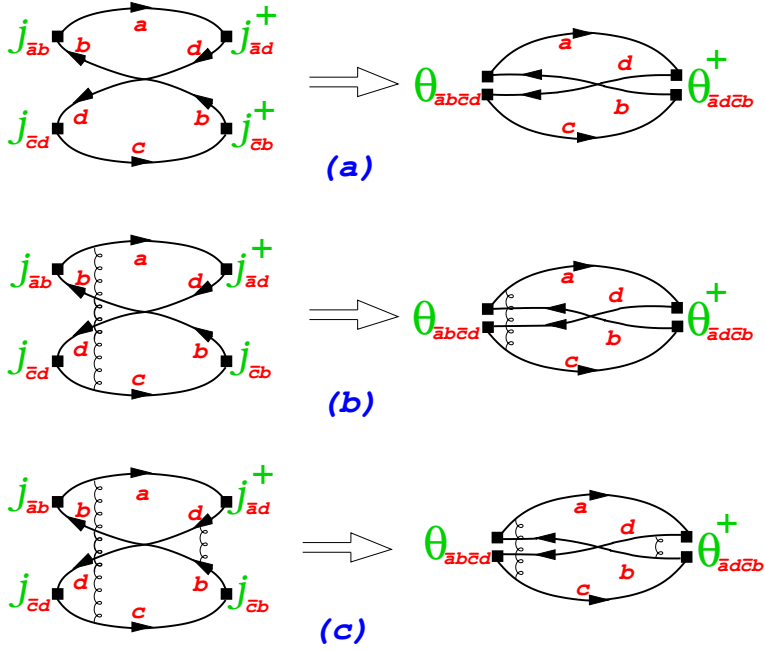}}\caption{Flavour-\emph{redistributing\/}
correlation function \cite{ESRp,ESRr}: illustrative examples of
contributions of the three (a,b,c) lowest perturbative orders and
their contractions to two tetraquark-interpolating
operators~$\theta$.}\label{Fr2}\end{figure}

\subsubsection{Two-Point Correlation Functions: Two Different
Tetraquark Interpolating Operators}\label{r2.}
\begin{figure}[htb]\centering{\includegraphics[scale=.401,clip]
{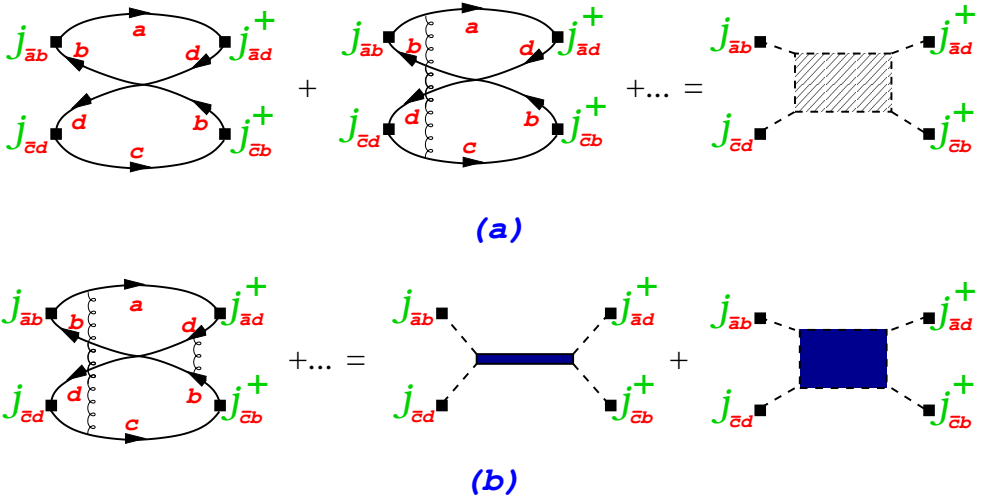}}\caption{Flavour-\emph{reordering\/}
correlation function of four quark-bilinear operators \cite{ESRr},
disintegrable into (a) an equation with no two-meson $s$-channel
cut (hatched rectangle) and (b) a tetraquark-adequate QCD sum rule
with two-meson $s$-channel cut (filled rectangle) and even
tetraquark poles (blue horizontal bar).}\label{Frg}\end{figure}

In the
flavour-reordering case \cite{ESRp,ESRr}, things get slightly more
complicated by the fact that~none of the QCD-level contributions
(Fig.~\ref{Fr2}) proves to be configuration-space separable. In
order to figure out which of these are tetraquark-phile, we indeed
have to exploit the Landau equations \cite{LDL}. These again point
to the necessity of two or more gluon exchanges of adequate
topology, thus opening the path for a systematics of contributions
relevant or not for tetraquarks (Fig.~\ref{Frg}).

\subsubsection{Three-Point Correlation Function -- One Tetraquark
and Two Conventional Mesons}\label{r3.}
\begin{figure}[htb]\centering{\includegraphics[scale=.401,clip]
{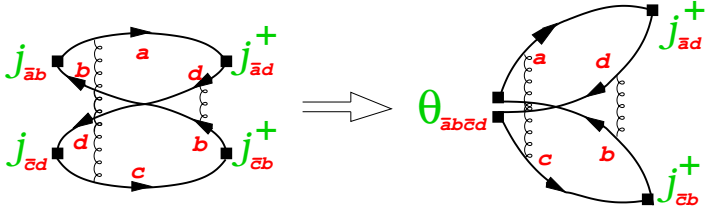}}\caption{Flavour-\emph{rearranging\/}
correlation function \cite{ESRr}: lowest-order
\emph{tetraquark-phile\/} contribution only once contracted to one
tetraquark-interpolating operator, leaving two quark-bilinear
currents untouched.}\label{Fr3}\end{figure}

Due to their common origin
in the four-point correlation function (\ref{4}) already noted,
the results gained for the two-point contraction evidently hold
for the three-point contraction (Fig.~\ref{Fr3}) too.

\section{Tetraquark-Adequate QCD Sum Rules: Flavour-Exotic
Tetraquarks}\label{S}Application of the prescription, indicated in
Sect.~\ref{M}, for the extraction of a QCD sum rule~from
correlation functions of hadron interpolating operators entails a
kind of ``canonical'' \emph{structure\/}:

Evaluation of the correlation function at QCD level necessitates
two sorts of contributions.\begin{itemize}\item On the one hand,
there exist perturbative contributions, discriminated by the
involvement of the strong coupling and usually cast in the form of
a dispersion integral of a spectral~density. In this integration,
the assumed cancellation of higher QCD vs.\ higher hadron
contributions is taken into account by an effective threshold
$s_{\rm eff}$ which by more in-depth analysis has been revealed to
depend (in a well-defined manner) on the Borel parameter $\tau$:
$s_{\rm eff}=s_{\rm eff}(\tau)$ \cite{ET1,ET2,ET3}.\item On the
other hand, there are nonperturbative contributions, favourably
subsumed in vacuum condensates -- vacuum expectation values of
products of the quark and gluon field operators of QCD -- which
may be understood as a kind of effective parameters of QCD. Upon a
Borel transformation, that is, in their Borel-transformed (or
``Borelized'', for short) form, these get multiplied by powers of
its Borel parameter, whence they are also called
\emph{power~corrections}.\end{itemize}For the set of
flavour-exotic tetraquarks, the lower-order perturbative QCD-level
contributions have been analyzed in Sect.~\ref{E}; an analogous --
albeit somewhat more involved -- discussion can be carried out,
with similar results, for the nonperturbative QCD-level
contributions~\cite{ESRp,ESRr,LMS10}.

Evaluation of this correlation function at the level of the
spectrum of hadron states leads~to expressions that involve -- in
addition to the Borel parameter $\tau$ -- (some of) the
hadron-relevant quantities of desire. In the case of any
flavour-exotic tetraquark state $|T\rangle$ specified by
Eq.~(\ref{F}), these observables include its mass $M$, its decay
constants $f$, and its vacuum--tetraquark matrix elements of a
pair of quark--antiquark bilinear operators (\ref{b}) in momentum
space, as defined~by\begin{equation}f_{\overline ab\overline cd}
\equiv\langle0|\theta_{\overline ab\overline cd}|T\rangle\ ,\qquad
f_{\overline ad\overline cb}\equiv\langle0|\theta_{\overline ad
\overline cb}|T\rangle\ ,\label{f}\end{equation}\begin{equation}
\begin{array}{l} \langle0|{\rm T}[j_{\overline ab}(y)\,j_{\overline
cd}(y')]|T\rangle\xrightarrow[\text{transformation}]{\text{Fourier}}
A(T\to j_{\overline ab}\,j_{\overline cd})\ ,\\[2ex]\langle0| {\rm
T}[j_{\overline ad}(y)\,j_{\overline cb}(y')]|T\rangle\xrightarrow
[\text{transformation}]{\text{Fourier}}A(T\to j_{\overline ad}\,
j_{\overline cb})\ .\end{array}\label{A}\end{equation}

By equating the outcomes of the above two levels of evaluation, we
eventually arrive at the generic shape of multiquark-adequate QCD
sum rules. It reads, for flavour-exotic~tetraquarks,\begin{align*}
(f_{\overline ab\overline cd})^2\exp(-M^2\,\tau)&=\int_{\hat
s}^{s_{\rm eff}(\tau)}\hspace{-2.58ex}{\rm d}s\exp(-s\,\tau)\,
\rho_{\rm p}(s)+\mbox{Borelized power corrections}\ ,\\
f_{\overline ab\overline cd}\,A(T\to j_{\overline ab}\,
j_{\overline cd})\exp(-M^2\,\tau)&=\int_{\hat s}^{s_{\rm
eff}(\tau)}\hspace{-2.58ex}{\rm d}s\exp(-s\,\tau)\,\Delta_{\rm
p}(s)+\mbox{Borelized power corrections}\ ,\\f_{\overline
ab\overline cd}\,f_{\overline ad\overline cb}\exp(-M^2\,\tau)&=
\int_{\hat s}^{s_{\rm eff}(\tau)}\hspace{-2.58ex}{\rm d}s
\exp(-s\,\tau)\,\rho_{\rm r}(s)+\mbox{Borelized power corrections}
\ ,\\f_{\overline ad\overline cb}\,A(T\to j_{\overline ab}\,
j_{\overline cd})\exp(-M^2\,\tau)&=\int_{\hat s}^{s_{\rm
eff}(\tau)}\hspace{-2.58ex}{\rm d}s\exp(-s\,\tau)\,\Delta_{\rm
r}(s)+\mbox{Borelized power corrections}\ ,\end{align*}wherein the
$s$-dependent functions $\rho_{\rm p,r}$ and $\Delta_{\rm p,r}$
denote, respectively, the flavour-preserving (p) and
flavour-rearranging (r) spectral densities in their
tetraquark-phile refinement explicated~in Sect.~\ref{T}, all
originating in the two-point ($\rho$) and three-point ($\Delta$)
correlation functions of Sect.~\ref{E}.

\vspace{7.422ex}
\begin{acknowledgement}\noindent{\bf Acknowledgements.~~~}Both
D.~M.\ and H.~S.\ would like to thank for support by joint
CNRS/RFBR Grant PRC Russia/19-52-15022, D.~M.\ for support by
Austrian Science Fund (FWF) Project P29028-N27, and H.~S.\ for
support by EU research and innovation program Horizon 2020 under
Grant Agreement~824093.\end{acknowledgement}

\nocite{*}
\bibliography{Zoom}

\begin{thebibliography}{21}

\bibitem{QSR}
M.A. Shifman, A.I. Vainshtein, V.I. Zakharov, Nucl.~Phys.~B \textbf{147}, 385
  (1979)

\bibitem{KGW}
K.G. Wilson, Phys.~Rev. \textbf{179}, 1499 (1969)

\bibitem{ET1}
W.~Lucha, D.~Melikhov, S.~Simula, Phys.~Rev.~D \textbf{79}, 096011 (2009)

\bibitem{ET2}
W.~Lucha, D.~Melikhov, S.~Simula, J.~Phys.~G \textbf{37}, 035003 (2010)

\bibitem{ET3}
W.~Lucha, D.~Melikhov, S.~Simula, Phys.~Lett.~B \textbf{687}, 48 (2010)

\bibitem{ESRp}
W.~Lucha, D.~Melikhov, H.~Sazdjian, Phys.~Rev.~D \textbf{100}, 014010 (2019)

\bibitem{ESRr}
W.~Lucha, D.~Melikhov, H.~Sazdjian, Phys.~Rev.~D \textbf{100}, 074029 (2019)

\bibitem{TMA1}
W.~Lucha, D.~Melikhov, H.~Sazdjian, PoS \textbf{(EPS-HEP2019)}, 536 (2020)

\bibitem{TMA2}
W.~Lucha, D.~Melikhov, H.~Sazdjian, EPJ Web Conf. \textbf{222}, 03016 (2019)

\bibitem{LMS10}
W.~Lucha, D.~Melikhov, H.~Sazdjian, Phys.~Rev.~D \textbf{103}, 014012 (2021)

\bibitem{TMA3}
W.~Lucha, D.~Melikhov, H.~Sazdjian, Supl.~Rev.~Mex.~F{\'i}s. \textbf{3},
  0308035 (2022)

\bibitem{LMS4}
W.~Lucha, D.~Melikhov, H.~Sazdjian, EPJ Web Conf. \textbf{192}, 00044 (2018)

\bibitem{PDG}
R.L. Workman et~al. (Particle Data Group), PTEP \textbf{2022}, 083C01 (2022)

\bibitem{RLJ}
R.L. Jaffe, Nucl.~Phys.~A \textbf{804}, 25 (2008)

\bibitem{MF}
M.~Fierz, Z.~Phys. \textbf{104}, 553 (1937)

\bibitem{LMS111}
W.~Lucha, D.~Melikhov, H.~Sazdjian, Prog.~Part.~Nucl.~Phys. \textbf{120},
  103867 (2021)

\bibitem{TQC1}
W.~Lucha, D.~Melikhov, H.~Sazdjian, Phys.~Rev.~D \textbf{96}, 014022 (2017)

\bibitem{TQC2}
W.~Lucha, D.~Melikhov, H.~Sazdjian, Eur.~Phys.~J.~C \textbf{77}, 866 (2017)

\bibitem{TQP1}
W.~Lucha, D.~Melikhov, H.~Sazdjian, PoS \textbf{(EPS-HEP 2017)}, 390 (2018)

\bibitem{TQP2}
W.~Lucha, D.~Melikhov, H.~Sazdjian, Phys.~Rev.~D \textbf{98}, 094011 (2018)

\bibitem{LDL}
L.D. Landau, Nucl.~Phys. \textbf{13}, 181 (1959)

\end{thebibliography}
\bibliographystyle{woc}
\end{document}